***Theoretical novel medical isotope production with deuterium-tritium fusion technology***

Lee J. Evitts[a]*, Philip W. Miller[b], Chiara Da Pieve[c,1], Andrew Turner[a], Stefano Borini[a]

[a]UKAEA (United Kingdom Atomic Energy Authority), Culham Campus, Abingdon, Oxfordshire, OX14 3DB, UK.

[b]Molecular Sciences Research Hub, White City Campus, Imperial College London, London, W12 0BZ, UK.

[c]The Institute of Cancer Research, 123 Old Brompton Rd, London, SW7 3RP, UK.

*Corresponding author: lee.evitts@ukaea.uk

## Abstract

**Background**: The emergence and growth of fusion technology enables investigative studies into its applications beyond typical power production facilities. This study seeks to determine the viability of medical isotope production with the neutrons produced in an example large fusion device. Using FISPACT-II (a nuclear inventory code) and a simulated fusion spectrum, the production yields of a significant number of potentially clinically relevant (both in use and novel) medical isotopes were calculated. Comparative calculations were also conducted against existing production routes.

**Results**: Depending on the neutron flux of the fusion device, it could be an ideal technology to produce alpha-emitters such as $^{212}$Bi/$^{212}$Pb, it may be able to contribute to the production of $^{99m}$Tc/$^{99}$Mo, and could offer an alternative route in the production a few Auger-emitting candidates. There is also a long list of beta-emitting nuclides where fusion technology may be best placed to produce over existing technologies including $^{67}$Cu, $^{90}$Y and $^{47}$Sc.

**Conclusions**: It is theoretically viable to produce existing and novel medical isotopes with fusion technology. However, a significant number of assumptions form the basis of this study which would need to be studied further for any particular nuclide of interest.

## 1. Introduction

In nuclear medicine, a radioactive nuclide (radionuclide or colloquially, medical isotope) is attached to a vector (often a pharmaceutical) which is then used in the diagnostic and therapeutic treatment of disease. The choice of radionuclide is dependent on the application. For novel candidates, the primary points to consider for preliminary screening for medical applications include (1):

(i) A reliable, economical, high purity and high production yield of the radionuclide (low GBq ranges are suitable for pre-clinical studies while high GBq quantities are required for clinical use).

(ii) A half-life that compatible with the diagnostic/therapeutic procedure needs and the biological half-life of the vector.

(iii) The energy of the emitted radiation is sufficient to ensure the delivery of enough dose to the tumour (in the case of therapeutic radionuclides) while avoiding normal tissues.

(iv) Existing and validated radiochemistry and commercially available chelators to speed up the clinical translation.

*(v)* Additional, advantageous, improved or alternative physical characteristics compared to already available radionuclides.

[1] Present address: Medicines Discovery Catapult, Block 35, Mereside, Alderley Park, Macclesfield, Cheshire, SK10 4ZF, UK.

Most existing medical isotopes are typically produced in either a nuclear fission research reactor or accelerator. In general, the choice of facility is dependent upon the product e.g. neutron-rich nuclides (which decay via β -) would be produced in a fission reactor or with a neutron source, while neutron-deficient nuclides (which decay via β+) would be produced in a cyclotron (2). The most clinically used nuclide, $^{99}$Mo with its isomeric daughter $^{99m}$Tc, is typically extracted as a fission product from an irradiated uranium target (3) though there are alternative production routes available e.g. via cyclotron (4). There have also been studies to examine the feasibility of alternative, novel routes e.g. the extraction of medical isotopes from existing nuclear waste (5). Each production mechanism has an associated set of pros and cons including the availability and cost of the relevant target material, development of purification processes, operational costs of a facility, etc.

In nuclear fission, a neutron is absorbed by a fissile material nucleus which splits into multiple nuclides with additional neutrons that feed the process. Alternatively, in a fusion device light nuclei are merged to form a heavier nucleus, which releases energy in the process. Power producing devices focus on the fusion of deuterium and tritium (heavy isotopes of hydrogen) due to the required plasma temperature, fuel availability, reaction cross-section and the amount of energy that is released. A deuterium and tritium fusion reaction (D-T) produces an alpha particle and a 14.1 MeV neutron. In power-producing fusion devices, the high-energy neutron can be used to breed additional tritium to sustain the reaction, and to generate heat or electricity. There are a variety of machine designs that utilize the D-T reaction, particularly within a plasma, including tokamaks and stellarators. Additionally, accelerator-based systems are available that could use D-T for other purposes, such as the production of medical isotopes. For example, SHINE Technologies will use D-T fusion in combination with a neutron multiplier (which are lower in energy) and a low-enriched uranium solution that will ideally produce Mo-99 at a sufficient scale (6).

There are existing experimental studies examining the production of medical isotopes with accelerators capable of delivering 14 MeV neutrons, particularly for the production of $^{99m}$Tc / $^{99}$Mo (7–9). This study examines a more general theoretical feasibility of utilizing available high-energy neutrons in a D-T machine (specifically a tokamak type, without a multiplier or enriched uranium present) to produce medical isotopes and how the yields compare to existing production routes.

## 2. Methodology

A list of potential radionuclides was produced by probing the chart of nuclides, filtering for those with a clinically suitable half-life (between one hour and twenty days, approximately corresponding to the range of existing medical radionuclides) and the condition that they decay to a stable daughter nuclide. Isomers (besides $^{99m}$Tc) were not considered in this study. For each potential radionuclide, their most likely nuclear reactions were identified such that the appropriate target material for each reaction could be determined. Very long-lived nuclides were also included as suitable target materials.

FISPACT-II (10) was used to perform batch nuclide inventory calculations for each product and potential reaction identified in the list. FISPACT-II is a code that can be used to model the activation and subsequent radioactive decay of a material during an irradiation period and subsequent cooling period. Calculations were performed using an example neutron spectrum of a fusion device and, depending on the nuclide, with either a cyclotron or fission research reactor to enable comparisons between different production routes. The High Flux Reactor (HFR) (11) is used for comparisons against a fission research reactor with an assumed high neutron flux of 5 x $10^{14}$ n/(cm$^2$ s), and an 11-12 MeV[2] proton beam at 100 µA is simulated for cyclotron comparisons. The TENDL-2019 nuclear data library was used for all neutron irradiation calculations as it provides a complete dataset for all nuclides and reaction channels, and the TENDL-2017 nuclear data library was used for proton irradiation calculations due to availability of data (12).

[2] An energy range is used due to the energy binning of FISPACT-II input spectra.

An example neutron spectrum of the outboard wall of a tokamak (whereby the plasma is confined in a torus through strong magnetic fields) device is generated from a D-T plasma source and OpenMC (13); an open-source neutron transport code with a simple tokamak style geometry generated by Paramak (14). The neutron spectra for both the fission research reactor and D-T fusion device are compared in Figure 1. The distribution of the fusion device contains a peak at 14.1 MeV from the D-T neutrons, with a lower energy distribution from scattered neutrons. The neutron flux is scaled to 5 x $10^{14}$ n/(cm$^2$ s) in the FISPACT-II calculations to enable an easier comparison with the fission results. However, as the neutron spectrum and flux is dependent on the type and scale of a device and position of the target material, more realistic calculations would be required for any particular configuration.

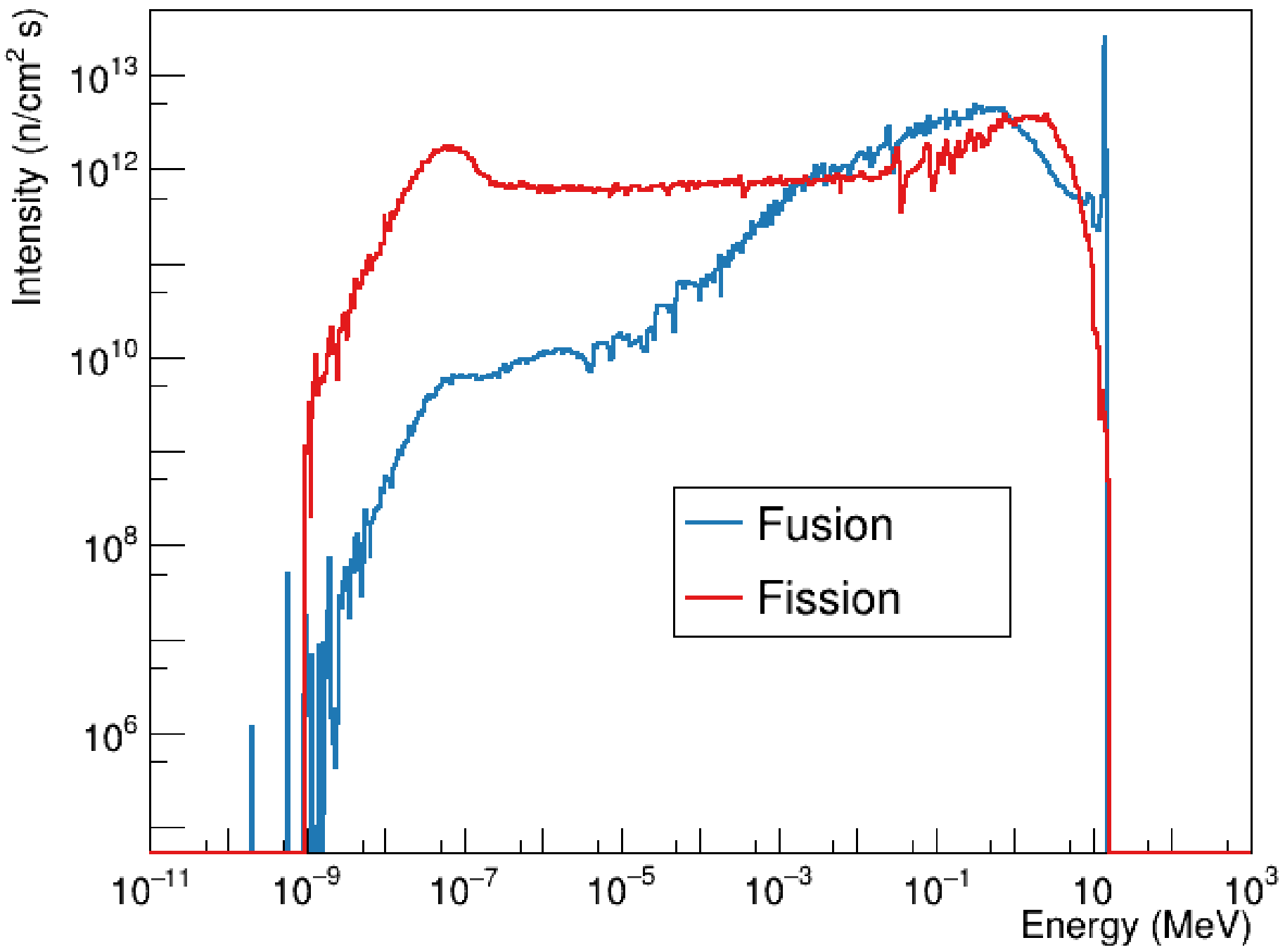

*Figure 1: Neutron spectra used in inventory calculations for (blue) a simulated first wall of a large D-T tokamak device, and (red) the High Flux Reactor at Petten with where both have an assumed total flux of 5 x $10^{14}$ n/cm$^2$ s.*

## 2.1 Assumptions

Due to the batch-wise nature of this study, a number of assumptions have to be made for each calculation, detailed below.

- Clinically relevant radionuclides have a half-life between one hour and twenty days, approximately corresponding to the range of existing medical radionuclides.
- The irradiated target is 1 g and enriched to be 100% isotopically pure. The quantity and purity of a real target will vary according to the material (e.g. some material may have to be suspended in a liquid or formed as an intermetallic), making direct comparisons between different irradiation methods and targets difficult.
- The irradiation period is three half-lives of the product with no upper bound i.e. does not take into consideration the operational schedule of a device.
- The neutron flux is 5 x $10^{14}$ n/(cm$^2$ s) for a fusion device, which may be optimistic of the first wall region of a large tokamak. This particular location would interfere with the tritium breeding but the results should be sufficient to demonstrate production feasibility. All results should be easily scaled according to the setup and the target mass.
- The product's element can be extracted from other elements with 100% efficiency, but the product cannot be separated from its isotopes. For example, Lu can be extracted from its neighbouring elements but $^{177}$Lu cannot be separated from $^{175}$Lu. There are techniques for isotope separation, but these would have to be studied on an individual basis if deemed necessary.

- The processes of extraction, purification, characterisation and quality control, pharmaceutical attachment, and shipping can take up to a day. The production yields for two time points (immediately after irradiation, and 24 hours after irradiation) are presented. The calculated production yields will be optimistic but should be suitable for comparisons against other production routes and as a point to initiate further research.

## 2.2 Accuracy of the simulations

The neutron flux and target mass are chosen such that results can easily be scaled to any machine. However, due to the assumptions used in this analysis, a discrepancy may arise between any validation experiment and the results shown here. These discrepancies could be due to the shape of the neutron spectrum (e.g. if there is a comparatively greater flux at higher energies), the presence of impurities in the target material, the length of the irradiation period, and the efficiency and time-scale of extracting the product following irradiation. There may also be a discrepancy arising from the choice of nuclear data library i.e. TENDL was chosen because it provides a complete dataset for all nuclides and reaction channels but it is based on a nuclear modelling system, where in reality there may be very limited experimental measurements of that reaction's cross section.

The results presented in the next chapter can be informative regarding which nuclides could be produced with good yields in a D-T fusion machine, but additional calculations should be conducted prior to any experimental campaign.

## 2.3 Reactions

Only (n, p), (n, α), (n, Xγ), (n, 2n) reactions are considered in this study, and any of these reactions which then decay into the desired product. For example, to produce $^{47}$Sc the $^{47}$Ti(n, p), $^{50}$V(n, α) and $^{45}$Sc(n, 2n) reactions are studied, along with any similar reactions that produce $^{47}$Ca, which β- decays into $^{47}$Sc. These reactions are chosen as their threshold energies are likely within range of the neutron spectrum shown in Figure 1 (i.e. below the upper limit of 14.1 MeV). There are other reactions available but these would typically be at a higher threshold energy or have a lower reaction cross-section. The (n, d) reaction is not included due to its generally higher reaction threshold energy and it would typically be dominated by the competing (n, p) reaction, given that one of the assumptions is that isotopes cannot be separated. For example, a list of all viable neutron production routes for $^{47}$Sc are listed in Table 1, along with the calculated production activities and radiochemical purities using the D-T fusion spectrum in Figure 1. The two most promising reactions, based on their production yields alone, would be with $^{50}$V and $^{48}$Ca targets. The production yield and purity for the $^{48}$Ti(n, d) reaction are both very low; the largest impurity being from the $^{48}$Ti(n, p)$^{48}$Sc reaction.

*Table 1: Calculated production yields and purities for all feasible $^{47}$Sc D-T neutron production routes.*

| Reaction | Approximate threshold energy (MeV) | Activity per target gram (GBq) | Radiochemical purity (%) |
|---|---|---|---|
| $^{47}$Ti(n, p)$^{47}$Sc | 0.1 | 132 | 97.5 |
| $^{48}$Ti(n, d)$^{47}$Sc | 10 | 4 | 7.5 |
| $^{50}$Ti(n, α)$^{47}$Ca -> $^{47}$Sc | 4 | 6.2 | 99.99 |
| $^{50}$V(n, α)$^{47}$Sc | 0.1 | 36.3 | 99.99 |
| $^{51}$V(n, n+α)$^{47}$Sc | 10 | <0.1 | <0.1 |
| $^{45}$Sc(2n, 2γ)$^{47}$Sc | 0 | <0.1 | <0.1 |
| $^{46}$Ca(n, γ)$^{47}$Ca -> $^{47}$Sc | 0 | 13.2 | 99.99 |
| $^{48}$Ca(n, 2n)$^{47}$Ca -> $^{47}$Sc | 10 | 615 | 99.99 |

## 3. Results & Discussion

The theoretical production yields calculated from FISPACT-II are summarized in the following sub-sections, categorized according to the type of radiation that is emitted in their decay. There are three properties calculated and used to compare with other production routes:

i) The molar activity, $A_m$, is the proportion of radiation emitted from the nuclide of interest to its molar mass (GBq/μmol). According to the assumptions detailed in Section 2.1, the product's element is extracted with 100% efficiency but not the individual isotopes. It is assumed that a molar activity above 200 GBq/μmol is suitable and, when comparing production routes, a higher value is better.

ii) The activity, $A$, is the total amount of activity from the nuclide of interest (GBq). As all calculations are performed on 1 g of target material, this can be approximately scaled according to feasible target amounts and neutron flux.

iii) The radiochemical purity, $P$, is the percentage of radiation emitted from the desired radionuclide with other isotopes in the extracted product. It is assumed that $P$ must be above 99% to have a suitable product.

There are two time periods in the calculations, at 0 hours (i.e. immediate) and 24 hours after irradiation where the product's element is extracted and the three production properties calculated. Only the most promising results from one of the two time periods are shown for clearer discussion.

### 3.1 Electron ($\beta^-$) emitters

The theoretical production yields for all physically suitable radionuclides (i.e. possessing a suitable half-life and decaying to stability) are listed for both fusion and fission devices in Table 4 in Appendix A. Note that the fission comparison only includes irradiation of similarly isotopically pure targets, and not the separation of products from fissile targets. Based on the production yields, the radionuclides that appear to be better suited for production with D-T fusion technology are summarized in Table 2.

*Table 2: Physical properties of selected β⁻ emitters and their production yields, where D-T fusion technology appears to be the best production route. The product (prod.) information includes the half-life ($T_{1/2}$), mean β- energy ($\bar{E}_{\beta-}$), γ energy ($E_\gamma$) of the most prominent emissions (including their branching ratios, BR), cooling period (cool.), natural abundance (nat. a.) of the target, molar activity (GBq/μmol) ($A_m$) and total activity (GBq) per target gram (A).*

| Prod. | $T_{1/2}$ | $\bar{E}_{\beta-}$ (keV) [BR] | $E_\gamma$ (MeV) [BR] | Cool. (h) | Target | Nat. A. (%) | $A_m$ | A |
|---|---|---|---|---|---|---|---|---|
| $^{24}$Na | 15 h | 555 [100%] | 1.37 [100%] | 0 | $^{24}$Mg | 79 | 3955 | 416 |
| | | | 2.75 [100%] | | $^{27}$Al | 100 | 6120 | 235 |
| $^{41}$Ar | 109.6 m | 459 [99%] | 1.29 [99%] | 0 | $^{41}$K | 6.7 | 6842 | 60 |
| | | | | | $^{44}$Ca | 2.1 | 59123 | 22 |
| $^{42}$K | 12.4 h | 824 [18%] | 1.52 [18%] | 0 | $^{42}$Ca | 0.65 | 5565 | 276 |
| | | 1566 [82%] | | | $^{45}$Sc | 100 | 8916 | 61 |
| $^{47}$Sc | 3.3 d | 143 [68%] | 0.16 [68%] | 24 | $^{50}$V | 0.25 | 1441 | 36 |
| | | 204 [32%] | | | $^{48}$Ca | 0.19 | 1443 | 615 |
| $^{48}$Sc | 43.7 h | 159 [10%] | 0.98 [100%] | 24 | $^{51}$V | 99.8 | 2657 | 10 |
| | | 227 [90%] | 1.04 [98%] | | | | | |
| | | | 1.31 [100%] | | | | | |
| $^{56}$Mn | 2.6 h | 255 [15%] | 0.85 [99%] | 0 | $^{56}$Fe | 92 | 24073 | 106 |
| | | 382 [28%] | 1.81 [27%] | | $^{59}$Co | 100 | 39462 | 27 |
| | | 1217 [57%] | 2.11 [14%] | | | | | |
| $^{65}$Ni | 2.5 h | 221 [28%] | 1.11 [15%] | 0 | $^{65}$Cu | 31 | 2205 | 13 |

| Prod. | $T_{1/2}$ | $\bar{E}_{\beta-}$ (keV) [BR] | $E_\gamma$ (MeV) [BR] | Cool. (h) | Target | Nat. A. (%) | $A_m$ | A |
|---|---|---|---|---|---|---|---|---|
| | | 372 [10%] | 1.48 [24%] | | $^{68}$Zn | 19 | 25557 | 14 |
| | | 875 [60%] | | | | | | |
| $^{67}$Cu | 61.8 h | 121 [57%] | 0.09 [16%] | 24 | $^{67}$Zn | 4 | 1774 | 20 |
| | | 154 [22%] | 0.18 [49%] | | | | | |
| | | 189 [20%] | | | | | | |
| $^{72}$Ga | 14.1 h | 219 [16%] | 0.63 [26%] | 0 | $^{72}$Ge | 27 | 1874 | 25 |
| | | 226 [22%] | 0.83 [95%] | | $^{75}$As | 100 | 6661 | 10 |
| | | 344 [29%] | 2.20 [27%] | | | | | |
| $^{76}$As | 1.1 d | 993 [35%] | 0.56 [45%] | 24 | $^{76}$Se | 9.2 | 1555 | 20 |
| | | 1264 [51%] | | | $^{79}$Br | 51 | 3774 | 4 |
| $^{78}$As | 90.7 m | 607 [16%] | 0.61 [54%] | 0 | $^{81}$Br | 49 | 64635 | 3.5 |
| | | 1228 [15%] | 0.69 [17%] | | | | | |
| | | 1560 [19%] | 1.31 [13%] | | | | | |
| | | 1956 [32%] | | | | | | |
| $^{82}$Br | 35.3 h | 138 [99%] | 0.55 [72%] | 24 | $^{82}$Kr | 12 | 2446 | 9 |
| | | | 0.62 [44%] | | $^{85}$Rb | 72 | 3253 | 2.2 |
| | | | 0.78 [84%] | | | | | |
| $^{87}$Kr | 76.3 m | 1502 [41%] | 0.40 [50%] | 0 | $^{87}$Rb | 28 | 45621 | 6.8 |
| | | 1694 [31%] | | | | | | |
| $^{86}$Rb | 18.6 d | 710 [91%] | 1.08 [9%] | 24 | $^{89}$Y | 100 | 259 | 4.2 |
| $^{90}$Y | 64 h | 932 [100%] | | 24 | $^{93}$Nb | 100 | 1063 | 4.7 |
| $^{109}$Pd | 13.7 h | 360 [100%] | 0.09 [4%] | 24 | $^{109}$Ag | 48 | 324 | 2.6 |
| $^{112}$Ag | 3.1 h | 1426 [20%] | 0.62 [43%] | 0 | $^{115}$In | 96 | 36404 | 1.1 |
| | | 1703 [54%] | | | | | | |
| $^{136}$Cs | 13 d | 99 [81%] | | 24 | $^{136}$Ba | 7.9 | 253 | 1.7 |
| | | 121 [14%] | | | $^{139}$La | 100 | 358 | 1.2 |
| $^{139}$Ba | 83 m | 941 [30%] | 0.17 [24%] | 0 | $^{142}$Ce | 11 | 64115 | 1.1 |
| | | 916 [70%] | | | $^{139}$La | 100 | 40747 | 1.3 |
| $^{140}$La | 1.7 d | 441 [11%] | 0.49 [46%] | 24 | $^{140}$Ce | 88 | 168 | 1.4 |
| | | 487 [44%] | 0.82 [23%] | | | | | |
| | | 629 [20%] | 1.60 [95%] | | | | | |
| $^{150}$Pm | 2.7 h | 499 [18%] | 0.33 [68%] | 0 | $^{150}$Sm | 7.4 | 36563 | 1.5 |
| | | 677 [19%] | 1.17 [16%] | | | | | |
| | | 895 [26%] | 1.32 [18%] | | | | | |
| $^{156}$Eu | 15 d | 146 [29%] | 0.09 [8%] | 24 | $^{156}$Gd | 20 | 148 | 2.0 |
| | | 965 [32%] | 0.81 [10%] | | $^{159}$Tb | 100 | 294 | 1.1 |
| $^{157}$Eu | 15 h | 296 [22%] | 0.06 [23%] | 0 | $^{157}$Gd | 16 | 7105 | 1.6 |
| | | 312 [15%] | 0.37 [11%] | | | | | |
| | | 462 [49%] | 0.41 [18%] | | | | | |
| $^{159}$Gd | 18.5 h | 189 [12%] | 0.36 [12%] | 0 | $^{159}$Tb | 100 | 4068 | 1.4 |
| | | 304 [29%] | | | | | | |
| | | 327 [59%] | | | | | | |
| $^{172}$Tm | 64 h | 668 [36%] | 0.08 [7%] | 24 | $^{172}$Yb | 22 | 1160 | 1.1 |
| | | 691 [29%] | | | | | | |
| $^{173}$Tm | 8.2 h | 272 [22%] | 0.40 [88%] | 0 | $^{173}$Yb | 16 | 13120 | 2.9 |

| Prod. | $T_{1/2}$ | $\bar{E}_{\beta-}$ (keV) [BR] | $E_\gamma$ (MeV) [BR] | Cool. (h) | Target | Nat. A. (%) | $A_m$ | A |
|---|---|---|---|---|---|---|---|---|
| | | 296 [76%] | | | | | | |
| $^{175}$Yb | 4.2 d | 19 [20%]<br>140 [73%] | 0.40 [13%] | 24 | $^{175}$Lu | 97 | 205 | 2.2 |
| $^{183}$Ta | 5.1 d | 190 [93%] | 0.11 [11%]<br>0.25 [27%]<br>0.35 [12%] | 24 | $^{183}$W | 14 | 821 | 1.1 |

The majority of nuclides in Table 2 are novel i.e. they have not been previously used for clinical purposes. Some exceptions include $^{67}$Cu and $^{90}$Y, the latter of which is typically used in radioembolization treatments of the liver (15). There are two existing routes to produce $^{90}$Y; the first is the direct neutron irradiation of $^{89}$Y to produce a carrier added form, and the second is from the decay of $^{90}$Sr, which is a uranium fission product with a long half-life (16). With a high molar activity and radiochemical purity, fusion may offer an alternate route in producing non-carrier added $^{90}$Y with a $^{93}$Nb target.

One example of a potentially clinically suitable novel nuclide that may be produced with D-T fusion is $^{47}$Sc. With a half-life of 3.3 days, it emits a β- with a mean energy of 162 keV, similar to other therapeutics like $^{177}$Lu (134 keV) and $^{131}$I (181 keV). There is also the emission of a 159 keV γ ray, which may be suitable for direct SPECT imaging, or could be coupled with the positron emitting $^{44}$Sc to form a theragnostic pair. As there is a similar coordination chemistry between $^{47}$Sc and $^{177}$Lu/$^{90}$Y, existing research on their chelation (e.g. with DOTA) can form a pre-existing basis of research (17). The two most viable reactions/targets identified with D-T fusion are $^{50}$V(n, α)$^{47}$Sc and $^{48}$Ca(n, 2n)$^{47}$Ca -> $^{47}$Sc which have threshold energies of approximately 100 keV and 10 MeV, respectively (12). However, both targets possess a very low natural abundance and would require purification processes to be developed.

Though fission may offer better production yields, it may also be possible to produce $^{31}$Si, $^{32}$P, $^{77}$As, $^{83}$Br, $^{105}$Rh, $^{111}$Ag, $^{131}$I, $^{149}$Pm, $^{161}$Tb, and $^{199}$Au in sufficient yields/purities with D-T fusion.

### 3.2 Alpha emitters

Suitable alpha-emitting radionuclides have been identified in previous studies (18,19), which form the list of investigated nuclides in this section. Inventory calculations were performed using three different targets composed of long-lived heavy nuclides ($^{229}$Th, $^{230}$Th, and $^{226}$Ra) but only a target of pure $^{226}$Ra yielded products with acceptable yields. In the study of beta-emitting nuclides, the irradiation time was linked with the product and the cooling time simulated for up to 24 hours. However, as there is only one target in this set of calculations with multiple products, the targets were irradiated for one week. The cooling period is also increased to two years, as some of the parent products in the decay chains can have a long half-life.

The production yields for four alpha-emitting radionuclides are shown in Figure 2. There is no perfect product with high values in each of the three categories, as the irradiation of the $^{226}$Ra target has a lot of products, each with long decay chains. However, $^{212}$Bi has a significant molar activity and it is very likely the radiochemical purity could be improved as it likely the longer-lived parent $^{212}$Pb would be extracted to form a generator, which is not considered here.

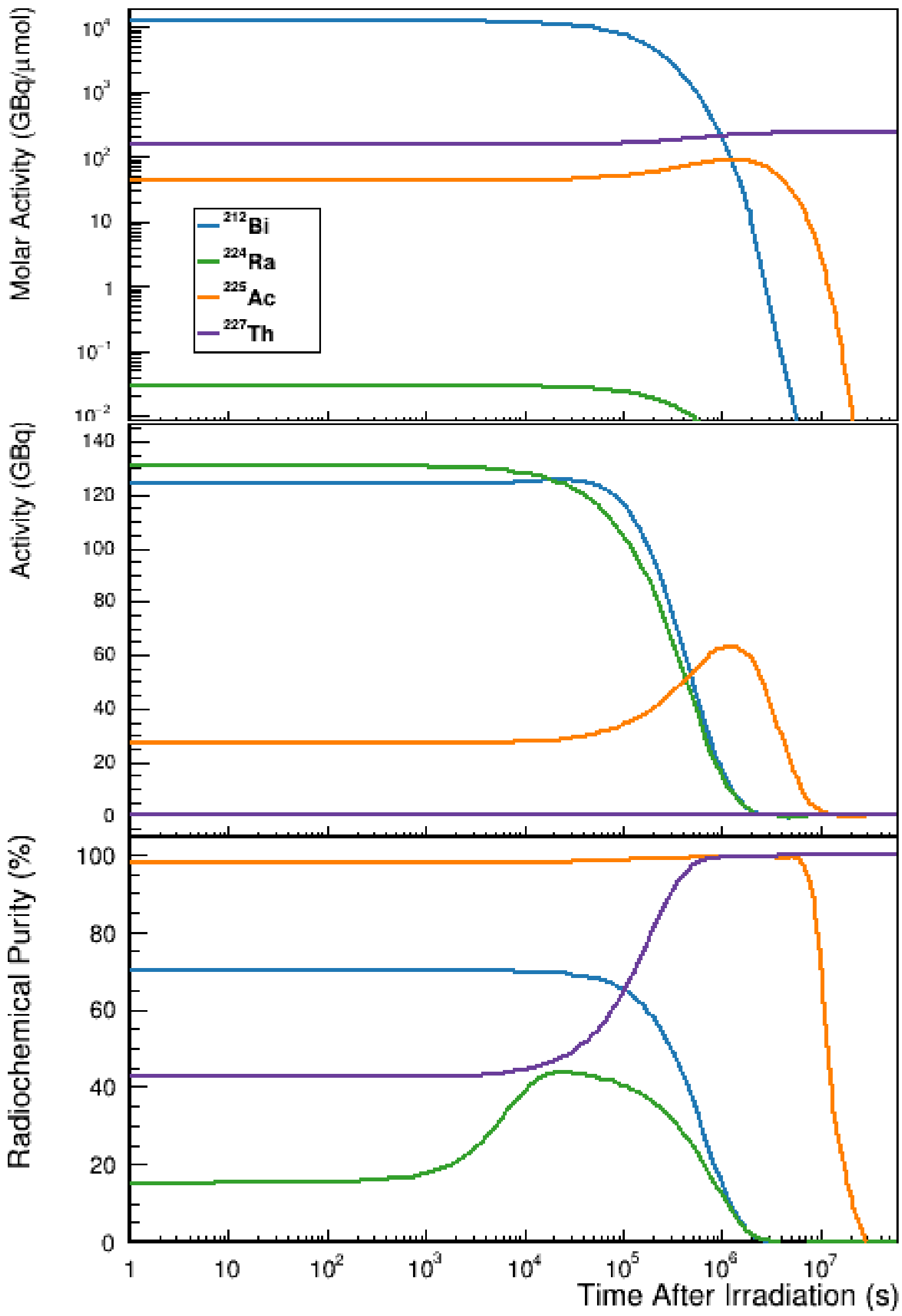

*Figure 2: Calculated production yields for four alpha-emitting radionuclides following irradiation with a D-T fusion device.*

### 3.3 Positron / Auger emitters

Several neutron-deficient nuclides may be produced with the available high-energy neutrons in a D-T fusion environment, as shown in Table 3, separated by the cooling period with the largest production purities and yields. The theoretical production yields are compared to a production route with an 11-12 MeV cyclotron and 100 µA beam current. Each product listed in Table 3 could be suitable for targeted therapies as they generally have a high intensity emission of low-energy Auger electrons or conversion electrons, except for $^{64}Cu$ which decays through both β- and β+ decay.

*Table 3: Theoretical production yields for positron/Auger emitters available through D-T fusion technology, compared with cyclotron production, 24 hours after irradiation. $A_m$ is the molar activity (GBq/µmol), A is the total activity (GBq) per target gram, and P is the radiochemical purity (%).*

| Prod. | $T_{1/2}$ | Simulated Fusion | Simulated Cyclotron |
|---|---|---|---|

| | | Target | $A_m$ | A | P | Target | $A_m$ | A | P |
|---|---|---|---|---|---|---|---|---|---|
| 0 hours after cooling: | | | | | | | | | |
| $^{64}$Cu | 12.7 h | $^{64}$Zn | 1180 | 171 | 99.99 | $^{64}$Ni | 5517 | 3155 | 99.99 |
| 24 hours after cooling: | | | | | | | | | |
| $^{77}$Br | 57 h | $^{78}$Kr | 248 | 331 | 100 | $^{77}$Se | 2034 | 1679 | 100 |
| $^{89}$Zr | 78 h | $^{92}$Mo | 338 | 11 | 99.91 | $^{89}$Y | 1471 | 2192 | 100 |
| $^{119}$Sb | 38.2 h | $^{120}$Te | 1725 | 304 | 99.58 | $^{119}$Sn | 3026 | 1074 | 99.98 |
| $^{135}$La | 19.5 h | $^{136}$Ce | 868 | 369 | 100 | $^{135}$Ba | 5949 | 690.8 | 100 |
| $^{201}$Tl | 3 d | $^{202}$Pb | 248 | 327 | 98.97 | $^{201}$Hg | 1112 | 162 | 45.7 |

The calculated production yields for $^{64}$Cu, $^{89}$Zr, $^{119}$Sb and $^{135}$La are theoretically higher in a cyclotron than a fusion device and their target materials are likely easier to acquire. However, fusion technology could offer a potential production route of these nuclides, depending on their relative accessibility and production costs with a cyclotron.

#### 3.3.1 Bromine-77

Bromine-77 has previously been considered for breast-tumour imaging (20) and a potential therapeutic Auger-emitter. However, production with a cyclotron is hindered due to the poor irradiation tolerance of Se. In this study, the calculations are always performed on a pure target and thus production yields may be overestimated for certain reactions e.g. in Ref (21), the production yields are improved in a cyclotron by using Co$^{77}$Se intermetallic targets. However, their molar activity of 700 GBq/µmol is significantly less than the pure form calculated here. Due to the inherent difficulties of production via a cyclotron it may be preferential to use fusion technology to produce $^{77}$Br although $^{78}$Kr has a low natural abundance and a gas target would introduce other challenges.

#### 3.3.2 Thallium-201

Thallium-201 decays via electron capture, emitting some intense low-energy Auger and conversion electrons along with a 167 keV γ ray in 10 % of decays (22). As such, $^{201}$Tl has previously been used for single photon emission tomography (23), and more recently considered for use in targeted Auger therapy (24,25). There are alternate production routes with a cyclotron e.g. the $^{203}$Tl(p, 3n)$^{201}$Pb → $^{201}$Tl that may offer improved production yields over the simple (p, n) route studied here though it would require a higher energy cyclotron. As such, it may also be preferential to use fusion technology to produce $^{201}$Tl though the target $^{202}$Pb is a very long-lived nuclide that may be difficult to acquire.

### 3.4 Molybdenum-99 / Technetium-99m

Based on the nuclear reactions discussed in Section 2.3, there are three potential targets to produce the longer-lived parent of $^{99m}$Tc in a fusion device ($^{98}$Mo, $^{100}$Mo and $^{102}$Ru). The calculations for simulating the production of $^{99m}$Tc are similar to those of previous sections except, immediately following the 8-day irradiation period, all Mo isotopes are extracted and their collective decay simulated for 24 hours. The production yields are then calculated under the same assumption that all Tc isotopes can be extracted, but not from each other.

With both the $^{98}$Mo and $^{102}$Ru targets, there is an insignificant amount of $^{101}$Tc after 24 hours of cooling as it is a short-lived nuclide (~14 minutes) but otherwise only $^{99m}$Tc and $^{99}$Tc are the only Tc isotopes that remain. In general, each of the three target materials produces $^{99m}$Tc with a molar activity of approximately 5400 GBq/µmol and a radiochemical purity above 99.999% but the total activity differs between targets. The best route is with a $^{100}$Mo target, producing 500 GBq per gram of target material, while the $^{98}$Mo and $^{102}$Ru targets produce 180 and 1 GBq per gram of target material, respectively.

Calculated comparisons are not made with a fission reactor as the typical route is via extraction from an irradiated uranium target which is likely a more productive route but, given the production yields calculated here, a fusion device may offer a solution when standard routes are inaccessible.

## 4. Conclusions

A significant number of assumptions have been made in this study, due to the inherent nature of batch-wise calculations, to inform the medical community of what fusion technology may be able to produce as it develops. One of the primary assumptions in this study is in defining a list of potential radionuclides based on their half-life and the stability of its daughter nuclide. However, not all radionuclides identified may be suitable for preclinical and clinical medical applications, and each nuclide of interest would require further refined studies. For example, longer lived radionuclides can have important logistic and production-related benefits e.g. manageable distribution and transport networks to distant centres and hospitals. However, after being injected in a patient, long-lived radionuclides have the time to wash out of the target organ, circulate and redistribute in non-target organs (e.g. clearance organs like the liver or kidneys) causing undesirable toxicity. Such damages could be mitigated when a short-half-life radionuclide is used instead.

Additional assumptions include the 100% efficient purification processes and the target material itself (both in its form, mass and purity), which may be both be challenging and costly but serve to draw comparisons in this study. One other key assumption is in the device itself; a large tokamak-type D-T fusion device with a significant neutron flux and availability (both in space and spare neutrons) that may also be a challenge to achieve. Results are given that can be scaled according to the flux of the targeted fusion device and the mass of the target, but additional calculations should be performed for any specific device and targeted nuclide. Such additional studies should also consider the geometry of the target itself as there may be routes to improve production yields.

This study also only lists the potential beta-emitters without going into detail on the energy and linear energy transfer (LET) of the emitted radiation, which are also very important aspects for the outcome of radionuclide therapy. The availability of a variety of alpha and beta-emitters with different energies and LETs could enable tailoring of the therapeutic effect to the therapeutic needs based on the target characteristics (e.g. location, size, geometry, and heterogeneity of the distribution of the molecular target) and minimise the damage to healthy tissue. Furthermore, an associated gamma-radiation emission could also be beneficial to verify, by SPECT imaging, the uptake pattern of the radiotherapeutic (based on its pharmacokinetics) and to calculate the absorbed doses to target tissue and normal organs. This would enable the delivery of an effective dose to the target and the minimising of side effects. Importantly, photon production, yield and energy, together with camera settings, are crucial for the image quality and dosimetry calculations. However, in the case of beta emitters, high energy gamma rays can affect the image resolution (i.e. production of blurred images) and are associated to high radiation exposure to operators and patients.

There is no consideration given to the *in vivo* stability of the potential radiopharmaceutical products and future studies would also need to consider their preparation, which depends on the chemical properties of each radioisotope. Radiolabelling of any molecule with radiohalogens (e.g. $^{131}I$ and $^{211}At$) requires the formation of a carbon-halogen bond while radiometals necessitate the presence of a suitable chelator. Unfortunately, a “one-chelator-fits-all” approach would be commercially and technically ideal (e.g. use of the same interchangeable scaffold with an imaging or therapeutic radionuclide) but realistically not always possible. In general, target sizes, target homogeneity, characteristics of the radionuclide, pharmacokinetics of the carrier molecule, administration route of the radiopharmaceutical, and expected risk to normal tissues can guide the selection of the radionuclide with the most suitable physical/chemical properties for the most favourable therapeutic effects.

In summary, there are many assumptions that feed into the calculated production yields in this study but the results should enable further discussions with the medical community. A significant number of beta emitters were identified that may be preferable to produce with fusion technology over accepted routes (e.g. $^{47}Sc$, $^{67}Cu$ and $^{90}Y$ among many). In addition, it was identified that the alpha-emitting $^{212}$Bi/$^{212}$Pb and a few potential Auger emitters may be produced in sufficient quantities. While a tokamak-type fusion machine was used in this study, it would also be feasible to produce many of the novel nuclides in an accelerator-based system without the enriched uranium and multiplier materials, potentially at a greater intensity depending on the reaction and beam intensities. For each nuclide of interest, additional studies would need to be performed including to (i) examine their nuclear reaction cross-section data, (ii) examine the cost of acquiring and recycling target materials, (iii) optimize the production yields through target design, (iv) develop target and product purification and extraction techniques, and (v) clinical research in e.g. *in vivo* stability, uptake and retention, etc.

# Declarations

**Funding sources**

Funding for this work was provided by the United Kingdom Atomic Energy Authority (UKAEA).

**Authors' contributions**

L.J.E. developed the methodology, conducted the modelling and analyzed the results. The manuscript was written by L.J.E. with contributions from C.D.P and P.W.M. A.T. cross-checked the methodology and the analysis. S.B. conceived the idea and led the project. All authors read and approved the final manuscript.

# Appendix A: Production yields for electron ($\beta^-$) emitters

*Table 4: Theoretical production yields for β− emitters with both D-T fusion technology and fission. $A_m$ is the molar activity (GBq/μmol), A is the total activity (GBq) per target gram, and P is the radiochemical purity (%).*

| Prod. | $T_{1/2}$ | Cool. (h) | Target | Fusion | | | Fission | | |
|---|---|---|---|---|---|---|---|---|---|
| | | | | $A_m$ | A | P | $A_m$ | A | P |
| $^{24}$Na | 15 h | 0 | $^{24}$Mg | 3955 | 416.1 | 100.00 | 7527 | 5.3 | 100.00 |
| | | | $^{27}$Al | 6120 | 234.6 | 100.00 | 7678 | 2.2 | 100.00 |
| | | | $^{23}$Na | trace | 44.3 | 99.85 | <0.1 | 856.8 | 100.00 |
| | | | $^{22}$Ne | trace | trace | 99.88 | <0.1 | trace | 100.00 |
| $^{31}$Si | 157.4 m | 0 | $^{31}$P | 2981 | 200.4 | 100.00 | 36335 | 91.5 | 100.00 |
| | | | $^{34}$S | 13545 | 184.1 | 100.00 | 41226 | 5.3 | 100.00 |
| | | | $^{30}$Si | trace | 30.1 | 100.00 | trace | 151.0 | 100.00 |
| $^{32}$P | 14.3 d | 24 | $^{32}$S | 91 | 470.0 | 100.00 | 337 | 160.2 | 99.99 |
| | | | $^{35}$Cl | 180 | 214.4 | 99.99 | 332 | 37.8 | 98.99 |
| | | | $^{31}$P | trace | 10.4 | 100.00 | trace | 192.2 | 100.00 |
| | | | $^{30}$Si | trace | trace | 100.00 | trace | trace | 100.00 |
| $^{41}$Ar | 109.6 m | 0 | $^{41}$K | 6842 | 59.9 | 100.00 | 37822 | 3.7 | 100.00 |
| | | | $^{40}$Ar | trace | 16.9 | 99.99 | <0.1 | 608.9 | 100.00 |
| | | | $^{44}$Ca | 59123 | 21.6 | 100.00 | 62661 | <0.1 | 100.00 |
| $^{42}$K | 12.4 h | 0 | $^{42}$Ca | 5565 | 275.5 | 100.00 | 9296 | 7.6 | 100.00 |
| | | | $^{45}$Sc | 8916 | 61.3 | 100.00 | 9353 | <1 | 99.99 |
| | | | $^{41}$K | trace | 190.4 | 100.00 | <0.1 | 1478.0 | 100.00 |
| | | | $^{40}$Ar | trace | trace | 100.00 | <0.1 | trace | 100.00 |
| $^{43}$K | 22.3 h | 24 | $^{43}$Ca | 5001 | 50.8 | 93.24 | 5224 | 1.9 | 99.92 |
| | | | $^{41}$K | trace | trace | 0.00 | trace | trace | 0.00 |
| $^{47}$Sc | 3.3 d | 24 | $^{47}$Ti | 889 | 131.8 | 97.52 | 1441 | 25.3 | 99.99 |
| | | | $^{50}$V | 1441 | 36.3 | 99.99 | 1442 | <1 | 99.96 |

| Prod. | $T_{1/2}$ | Cool. (h) | Target | Fusion | | | Fission | | |
|---|---|---|---|---|---|---|---|---|---|
| | | | | $A_m$ | A | P | $A_m$ | A | P |
| | | | $^{45}$Sc | trace | <0.1 | 0.01 | trace | 6.6 | 0.33 |
| | | | $^{48}$Ca | 1443 | 614.9 | 100.00 | 1443 | <1 | 99.99 |
| $^{48}$Sc | 43.7 h | 24 | $^{48}$Ti | 2310 | 45.7 | 93.43 | 2617 | <1 | 99.62 |
| | | | $^{51}$V | 2657 | 10.2 | 99.99 | 2656 | <0.1 | 99.99 |
| $^{56}$Mn | 2.6 h | 0 | $^{56}$Fe | 24073 | 105.7 | 100.00 | 43992 | 1.7 | 100.00 |
| | | | $^{59}$Co | 39462 | 27.3 | 100.00 | 44623 | <1 | 99.99 |
| | | | $^{55}$Mn | <0.1 | 442.8 | 99.88 | <1 | 9951.0 | 100.00 |
| | | | $^{54}$Cr | <0.1 | trace | 99.91 | <1 | trace | 100.00 |
| $^{61}$Co | 1.6 h | 0 | $^{61}$Ni | 42219 | 79.1 | 88.53 | 69895 | 3.1 | 99.88 |
| | | | $^{64}$Ni | 70316 | 3.2 | 39.24 | 70304 | trace | 39.01 |
| | | | $^{59}$Co | trace | trace | 0.00 | trace | <1 | 0.00 |
| $^{65}$Ni | 2.5 h | 0 | $^{65}$Cu | 2205 | 12.8 | 100.00 | 31205 | <1 | 100.00 |
| | | | $^{64}$Ni | trace | 49.6 | 99.99 | <0.1 | 868.1 | 100.00 |
| | | | $^{68}$Zn | 25557 | 14.0 | 100.00 | 43945 | <1 | 100.00 |
| $^{67}$Cu | 61.8 h | 24 | $^{67}$Zn | 1774 | 20.2 | 100.00 | 1872 | <1 | 100.00 |
| | | | $^{68}$Zn | 110 | 2 | 99.99 | | | |
| | | | $^{65}$Cu | trace | trace | 0.00 | trace | trace | 3.03 |
| $^{72}$Ga | 14.1 h | 0 | $^{72}$Ge | 1874 | 25.1 | 99.99 | 4246 | <1 | 99.99 |
| | | | $^{71}$Ga | <0.1 | 693.5 | 46.29 | <1 | 5598.0 | 99.97 |
| | | | $^{75}$As | 6661 | 9.6 | 100.00 | 7921 | <0.1 | 99.97 |
| $^{73}$Ga | 4.9 h | 0 | $^{73}$Ge | 18775 | 18.9 | 93.60 | 23426 | <1 | 99.52 |
| | | | $^{76}$Ge | 23773 | <1 | 36.29 | 23780 | trace | 41.06 |
| | | | $^{71}$Ga | trace | trace | 0.00 | trace | <1 | 0.01 |
| $^{75}$Ge | 82.8 m | 0 | $^{76}$Ge | <0.1 | 763.6 | 52.64 | trace | 1.0 | 0.59 |
| | | | $^{75}$As | 17425 | 15.1 | 58.39 | 67163 | <1 | 66.71 |
| | | | $^{74}$Ge | trace | 127.6 | 49.08 | <0.1 | 333.8 | 77.51 |
| | | | $^{78}$Se | 80851 | 3.4 | 56.27 | 82829 | trace | 60.45 |
| $^{76}$As | 1.1 d | 24 | $^{76}$Se | 1555 | 20.2 | 99.99 | 4170 | <1 | 99.82 |
| | | | $^{75}$As | <0.1 | 813.9 | 90.61 | <1 | 4885.0 | 99.97 |
| | | | $^{79}$Br | 3774 | 4.2 | 100.00 | 4319 | <0.1 | 99.97 |
| $^{77}$As | 38.8 h | 24 | $^{77}$Se | 2781 | 21.3 | 90.81 | 2983 | <1 | 99.80 |
| | | | $^{76}$Ge | 128 | 79.4 | 99.96 | 2954 | 207.3 | 100.00 |
| | | | $^{75}$As | trace | <0.1 | 0.00 | trace | 1.8 | 0.03 |
| $^{78}$As | 90.7 m | 0 | $^{78}$Se | 2623 | 17.1 | 93.96 | 2766 | <0.1 | 98.52 |
| | | | $^{81}$Br | 64635 | 3.5 | 99.26 | 70567 | trace | 99.64 |
| | | | $^{76}$Ge | trace | trace | 0.01 | 2 | <1 | 0.06 |
| $^{82}$Br | 35.3 h | 24 | $^{82}$Kr | 2446 | 9.2 | 100.00 | 3168 | <0.1 | 100.00 |
| | | | $^{81}$Br | <0.1 | 584.6 | 96.59 | <1 | 3681.0 | 100.00 |
| | | | $^{85}$Rb | 3253 | 2.2 | 100.00 | 3268 | trace | 100.00 |
| $^{83}$Br | 2.4 h | 0 | $^{83}$Kr | 43755 | 20.6 | 97.03 | 48029 | <1 | 99.81 |
| | | | $^{82}$Se | 445 | 13.0 | 99.97 | 44756 | 26.1 | 100.00 |
| | | | $^{81}$Br | trace | trace | 0.00 | trace | <1 | 0.00 |
| $^{87}$Kr | 76.3 m | 0 | $^{87}$Rb | 45621 | 6.8 | 100.00 | 65731 | trace | 99.99 |
| | | | $^{86}$Kr | trace | 5.1 | 5.49 | trace | 4.2 | 98.41 |
| $^{86}$Rb | 18.6 d | 24 | $^{87}$Rb | <0.1 | 624.4 | 100.00 | trace | <1 | 100.00 |

| Prod. | $T_{1/2}$ | Cool. (h) | Target | Fusion | | | Fission | | |
|---|---|---|---|---|---|---|---|---|---|
| | | | | $A_m$ | A | P | $A_m$ | A | P |
| | | | $^{86}Sr$ | 18 | 22.7 | 99.90 | 20 | <1 | 100.00 |
| | | | $^{85}Rb$ | <0.1 | 583.1 | 56.76 | <0.1 | 948.1 | 99.97 |
| | | | $^{89}Y$ | 259 | 4.2 | 100.00 | 256 | trace | 100.00 |
| $^{90}Y$ | 64 h | 24 | $^{90}Zr$ | 44 | 15.3 | 5.75 | 817 | <1 | 69.99 |
| | | | $^{89}Y$ | trace | 35.7 | 56.86 | <0.1 | 425.0 | 99.99 |
| | | | $^{93}Nb$ | 1063 | 4.7 | 99.68 | 1741 | <0.1 | 99.68 |
| $^{92}Y$ | 3.5 h | 0 | $^{92}Zr$ | 26440 | 12.8 | 94.74 | 30518 | <0.1 | 98.56 |
| $^{96}Nb$ | 23.4 h | 0 | $^{96}Mo$ | 3353 | 10.3 | 96.97 | 4533 | <0.1 | 99.36 |
| $^{97}Nb$ | 72.1 m | 0 | $^{97}Mo$ | 83195 | 10.4 | 74.82 | 93239 | <0.1 | 68.14 |
| $^{97}Zr$ | 16.7 h | 0 | $^{96}Zr$ | <0.1 | 126.2 | 85.49 | <0.1 | 502.3 | 99.98 |
| $^{105}Rh$ | 35.4 h | 24 | $^{105}Pd$ | 3267 | 22.6 | 99.99 | 3279 | <1 | 99.99 |
| | | | $^{104}Ru$ | 2558 | 301.1 | 83.71 | 3280 | 500.1 | 98.77 |
| | | | $^{103}Rh$ | trace | trace | 0.01 | trace | <1 | 85.13 |
| $^{109}Pd$ | 13.7 h | 24 | $^{110}Pd$ | <0.1 | 220.7 | 99.84 | trace | <1 | 45.13 |
| | | | $^{108}Pd$ | <0.1 | 430.4 | 100.00 | <1 | 6361.0 | 100.00 |
| | | | $^{109}Ag$ | 324 | 2.6 | 100.00 | <1 | <0.1 | 99.98 |
| | | | $^{112}Cd$ | 7866 | <1 | 100.00 | 8185 | trace | 100.00 |
| $^{111}Ag$ | 7.5 d | 24 | $^{111}Cd$ | 621 | 12.5 | 99.84 | 645 | <0.1 | 99.93 |
| | | | $^{110}Pd$ | 81 | 245.6 | 49.33 | 641 | 327.7 | 99.70 |
| $^{112}Ag$ | 3.1 h | 0 | $^{112}Cd$ | 31118 | 5.9 | 89.56 | 31461 | trace | 88.27 |
| | | | $^{115}In$ | 36404 | 1.1 | 99.20 | 36476 | trace | 99.27 |
| $^{113}Ag$ | 5.4 h | 0 | $^{113}Cd$ | 21104 | 6.6 | 48.87 | 21383 | <0.1 | 55.34 |
| | | | $^{116}Cd$ | 19940 | <0.1 | 5.75 | 19787 | trace | 5.61 |
| $^{115}Cd$ | 53.5 h | 24 | $^{116}Cd$ | <0.1 | 216.1 | 82.68 | trace | <1 | 49.94 |
| | | | $^{114}Cd$ | <0.1 | 197.5 | 98.02 | <0.1 | 691.9 | 98.73 |
| | | | $^{115}In$ | 79 | <1 | 63.73 | 185 | trace | 73.47 |
| | | | $^{118}Sn$ | 457 | <1 | 84.33 | 520 | trace | 86.47 |
| $^{121}Sn$ | 27 h | 24 | $^{122}Sn$ | <0.1 | 120.4 | 99.81 | trace | <1 | 65.63 |
| | | | $^{120}Sn$ | trace | 38.7 | 94.24 | trace | 68.2 | 99.99 |
| | | | $^{121}Sb$ | 5 | 1.4 | 99.91 | <1 | <0.1 | 99.96 |
| | | | $^{124}Te$ | 492 | <1 | 99.95 | 626 | trace | 99.96 |
| $^{126}Sb$ | 12.4 d | 24 | $^{126}Te$ | 103 | <1 | 98.78 | 111 | trace | 98.81 |
| $^{128}Sb$ | 9.1 h | 24 | $^{128}Te$ | 606 | <0.1 | 92.97 | 350 | trace | 93.37 |
| $^{127}Te$ | 9.4 h | 0 | $^{128}Te$ | <0.1 | 240.3 | 74.93 | trace | <1 | 0.35 |
| | | | $^{126}Te$ | <0.1 | 142.9 | 95.43 | <0.1 | 558.5 | 99.91 |
| | | | $^{127}I$ | 87 | 2.1 | 98.32 | <1 | <0.1 | 83.69 |
| | | | $^{130}Xe$ | 2705 | <1 | 98.75 | 3303 | trace | 99.07 |
| $^{126}I$ | 12.9 d | 24 | $^{127}I$ | <0.1 | 540.3 | 99.99 | trace | <1 | 99.90 |
| | | | $^{126}Xe$ | 2 | 8.0 | 3.49 | <0.1 | <0.1 | 31.28 |
| $^{130}I$ | 12.4 h | 24 | $^{130}Xe$ | 3870 | <1 | 100.00 | 6834 | trace | 99.99 |
| | | | $^{133}Cs$ | 8293 | <1 | 100.00 | 8687 | trace | 99.87 |
| $^{131}I$ | 8 d | 24 | $^{131}Xe$ | 594 | 2.4 | 98.75 | 600 | trace | 99.75 |
| | | | $^{130}Te$ | 26 | 36.0 | 99.90 | 584 | 80.8 | 100.00 |
| $^{132}I$ | 2.3 h | 0 | $^{132}Xe$ | 38550 | 1.5 | 88.90 | 41081 | trace | 91.20 |
| $^{133}Xe$ | 5.2 d | 24 | $^{134}Xe$ | <0.1 | 533.1 | 68.75 | trace | 1.2 | 33.83 |

| Prod. | $T_{1/2}$ | Cool. (h) | Target | Fusion | | | Fission | | |
|---|---|---|---|---|---|---|---|---|---|
| | | | | $A_m$ | A | P | $A_m$ | A | P |
| | | | $^{132}$Xe | <0.1 | 114.3 | 30.94 | <0.1 | 411.7 | 92.13 |
| | | | $^{133}$Cs | 8 | 6.0 | 64.01 | 13 | <0.1 | 68.13 |
| | | | $^{136}$Ba | 752 | <1 | 67.65 | 731 | trace | 70.43 |
| $^{136}$Cs | 13 d | 24 | $^{136}$Ba | 253 | 1.7 | 100.00 | 301 | trace | 100.00 |
| | | | $^{139}$La | 358 | 1.2 | 100.00 | 360 | trace | 100.00 |
| $^{139}$Ba | 83 m | 0 | $^{142}$Ce | 64115 | 1.1 | 100.00 | 69515 | trace | 100.00 |
| | | | $^{138}$Ba | trace | 10.4 | 2.94 | <0.1 | 115.3 | 99.60 |
| | | | $^{139}$La | 40747 | 1.3 | 99.66 | 51070 | trace | 99.83 |
| $^{140}$La | 1.7 d | 24 | $^{140}$Ce | 168 | 1.4 | 100.00 | 143 | trace | 100.00 |
| | | | $^{139}$La | <0.1 | 75.2 | 100.00 | <1 | 1915.0 | 100.00 |
| $^{142}$La | 91 m | 0 | $^{142}$Ce | 31362 | 1.2 | 94.93 | 18980 | trace | 96.13 |
| $^{142}$Pr | 19.1 h | 24 | $^{142}$Nd | 8 | 1.6 | 99.98 | 17 | trace | 99.97 |
| | | | $^{141}$Pr | <0.1 | 121.7 | 100.00 | <1 | 1535.0 | 99.98 |
| $^{143}$Pr | 13.6 d | 24 | $^{143}$Nd | 305 | 3.6 | 95.81 | 335 | trace | 99.07 |
| | | | $^{142}$Ce | 22 | 33.3 | 99.91 | 352 | 271.9 | 100.00 |
| $^{145}$Pr | 6 h | 0 | $^{145}$Nd | 18209 | 2.4 | 88.69 | 18934 | trace | 95.35 |
| | | | $^{148}$Nd | 19253 | <1 | 54.70 | 19305 | trace | 67.40 |
| $^{148}$Pm | 5.4 d | 24 | $^{148}$Sm | 303 | 1.8 | 85.99 | 422 | trace | 69.06 |
| | | | $^{151}$Eu | 274 | 1.7 | 83.58 | 13 | trace | 0.59 |
| $^{149}$Pm | 53 h | 24 | $^{148}$Nd | 832 | 206.7 | 99.61 | 2181 | 1095.0 | 99.99 |
| | | | $^{150}$Nd | 1837 | 474.5 | 74.12 | 15 | 2.6 | 0.38 |
| | | | $^{149}$Sm | 1743 | 2.6 | 96.10 | 1420 | trace | 97.47 |
| $^{150}$Pm | 2.7 h | 0 | $^{150}$Sm | 36563 | 1.5 | 99.16 | 37437 | trace | 99.35 |
| | | | $^{153}$Eu | 37327 | <1 | 99.20 | 40698 | trace | 99.67 |
| $^{153}$Sm | 46 h | 24 | $^{154}$Sm | <0.1 | 449.9 | 100.00 | trace | 2.2 | 99.51 |
| | | | $^{152}$Sm | <1 | 3052.0 | 100.00 | 21 | 132700.0 | 100.00 |
| | | | $^{156}$Gd | 2250 | <1 | 100.00 | 2331 | trace | 100.00 |
| $^{156}$Eu | 15 d | 24 | $^{156}$Gd | 148 | 2.0 | 99.75 | 144 | trace | 99.14 |
| | | | $^{159}$Tb | 294 | 1.1 | 99.79 | 314 | trace | 98.74 |
| $^{157}$Eu | 15 h | 0 | $^{157}$Gd | 7105 | 1.6 | 99.49 | 4990 | trace | 59.61 |
| $^{159}$Gd | 18.5 h | 0 | $^{160}$Gd | <1 | 636.7 | 71.00 | trace | 4.4 | 0.53 |
| | | | $^{158}$Gd | <0.1 | 550.1 | 100.00 | <1 | 3939.0 | 100.00 |
| | | | $^{159}$Tb | 4068 | 1.4 | 100.00 | 5038 | trace | 100.00 |
| | | | $^{162}$Dy | 5966 | <1 | 100.00 | 5272 | trace | 100.00 |
| $^{161}$Tb | 6.9 d | 24 | $^{161}$Dy | 592 | 2.2 | 98.52 | 662 | trace | 99.66 |
| | | | $^{160}$Gd | 86 | 205.6 | 99.96 | 691 | 655.9 | 100.00 |
| $^{165}$Dy | 2.3 h | 0 | $^{165}$Ho | 167 | 1.9 | 80.12 | 95 | trace | 56.73 |
| | | | $^{164}$Dy | <1 | 2329.0 | 58.07 | 85 | 515000.0 | 58.48 |
| | | | $^{168}$Er | 46135 | <1 | 78.28 | 49383 | <0.1 | 99.63 |
| $^{166}$Ho | 26.8 h | 24 | $^{166}$Er | 2 | <1 | 100.00 | 2 | trace | 99.89 |
| | | | $^{165}$Ho | <1 | 1439.0 | 100.00 | 4 | 21350.0 | 99.93 |
| | | | $^{169}$Tm | 1019 | <1 | 100.00 | 3246 | trace | 99.93 |
| $^{167}$Ho | 3 h | 0 | $^{167}$Er | 34034 | <1 | 99.57 | 29680 | trace | 99.64 |
| | | | $^{170}$Er | 35538 | <1 | 21.53 | 35787 | trace | 24.15 |
| | | | $^{165}$Ho | trace | <1 | 0.02 | <0.1 | 405.9 | 4.00 |

| Prod. | $T_{1/2}$ | Cool. (h) | Target | Fusion | | | Fission | | |
|---|---|---|---|---|---|---|---|---|---|
| | | | | $A_m$ | A | P | $A_m$ | A | P |
| $^{169}$Er | 9.4 d | 24 | $^{170}$Er | <0.1 | 573.5 | 92.88 | trace | 4.2 | 0.86 |
| | | | $^{168}$Er | <0.1 | 462.3 | 100.00 | <1 | 2391.0 | 100.00 |
| $^{172}$Tm | 64 h | 24 | $^{172}$Yb | 1160 | 1.1 | 99.89 | 1101 | trace | 99.84 |
| | | | $^{175}$Lu | 1673 | <1 | 99.96 | 1819 | trace | 99.91 |
| $^{173}$Tm | 8.2 h | 0 | $^{173}$Yb | 13120 | 2.9 | 99.20 | 13631 | trace | 99.59 |
| | | | $^{176}$Lu | 13537 | <1 | 99.47 | 13924 | trace | 99.02 |
| | | | $^{176}$Yb | 13794 | <1 | 26.69 | 14012 | trace | 64.25 |
| $^{175}$Yb | 4.2 d | 24 | $^{176}$Yb | <0.1 | 483.2 | 99.99 | trace | 5.4 | 96.84 |
| | | | $^{174}$Yb | <0.1 | 201.1 | 100.00 | 2 | 11090.0 | 100.00 |
| | | | $^{175}$Lu | 205 | 2.2 | 100.00 | <0.1 | trace | 99.96 |
| | | | $^{178}$Hf | 999 | <1 | 100.00 | 1047 | trace | 100.00 |
| $^{177}$Lu | 6.6 d | 24 | $^{177}$Hf | 505 | 2.1 | 98.74 | 476 | trace | 98.61 |
| | | | $^{176}$Lu | 1 | 6869.0 | 99.65 | 129 | 517600.0 | 99.98 |
| | | | $^{176}$Yb | 80 | 123.3 | 99.99 | 715 | 686.6 | 100.00 |
| | | | $^{180}$Hf | 727 | <1 | 99.58 | 727 | trace | 99.98 |
| | | | $^{175}$Lu | trace | 1.3 | 2.21 | <1 | 1479.0 | 82.95 |
| $^{179}$Lu | 4.6 h | 0 | $^{179}$Hf | 24573 | 1.5 | 94.89 | 24932 | trace | 97.57 |
| $^{183}$Ta | 5.1 d | 24 | $^{183}$W | 821 | 1.1 | 99.51 | 778 | trace | 99.63 |
| | | | $^{186}$W | 948 | <0.1 | 99.89 | 948 | trace | 99.35 |
| | | | $^{181}$Ta | trace | 9.5 | 3.63 | 3 | 16620.0 | 91.11 |
| $^{184}$Ta | 8.7 h | 0 | $^{184}$W | 11729 | <1 | 98.95 | 11959 | trace | 99.04 |
| | | | $^{187}$Re | 12948 | <1 | 99.78 | 13023 | trace | 99.71 |
| $^{187}$W | 24 h | 24 | $^{187}$Re | 68 | <1 | 99.90 | 10 | trace | 99.90 |
| | | | $^{186}$W | <0.1 | 348.1 | 95.28 | 2 | 12830.0 | 99.99 |
| | | | $^{190}$Os | 4547 | <1 | 100.00 | 4808 | trace | 99.99 |
| $^{186}$Re | 3.7 d | 24 | $^{187}$Re | <0.1 | 463.7 | 33.65 | trace | 3.2 | 0.03 |
| | | | $^{185}$Re | <1 | 3092.0 | 96.37 | 14 | 73340.0 | 100.00 |
| | | | $^{186}$Os | 19 | <1 | 99.94 | 12 | trace | 99.99 |
| $^{188}$Re | 17 h | 24 | $^{188}$Os | 4701 | <1 | 99.93 | 2822 | trace | 99.84 |
| | | | $^{187}$Re | <1 | 829.7 | 82.71 | 2 | 9794.0 | 99.98 |
| | | | $^{191}$Ir | 5714 | <0.1 | 99.97 | 5976 | trace | 98.48 |
| $^{189}$Re | 24 h | 24 | $^{189}$Os | 4509 | <1 | 97.10 | 4567 | trace | 98.62 |
| | | | $^{192}$Os | 4768 | trace | 99.95 | 4769 | trace | 99.95 |
| | | | $^{187}$Re | trace | <0.1 | 0.00 | trace | 1.5 | 0.01 |
| $^{191}$Os | 15.4 d | 24 | $^{192}$Os | <1 | 527.6 | 77.10 | trace | 2.6 | 0.55 |
| | | | $^{190}$Os | <1 | 551.0 | 75.99 | <1 | 3240.0 | 77.37 |
| | | | $^{191}$Ir | <1 | 1.9 | 80.36 | trace | trace | 0.78 |
| | | | $^{194}$Pt | 305 | <1 | 96.03 | 243 | trace | 99.31 |
| $^{193}$Os | 30.1 h | 24 | $^{193}$Ir | 669 | <1 | 99.65 | 174 | trace | 99.67 |
| | | | $^{192}$Os | <0.1 | 69.2 | 28.99 | <0.1 | 414.9 | 99.77 |
| | | | $^{196}$Pt | 3775 | <0.1 | 100.00 | 3748 | trace | 100.00 |
| $^{194}$Ir | 19.3 h | 24 | $^{194}$Pt | 3917 | <1 | 91.24 | 1817 | trace | 61.10 |
| | | | $^{193}$Ir | <1 | 1048.0 | 96.58 | 6 | 30650.0 | 99.90 |
| | | | $^{197}$Au | 5807 | <0.1 | 99.85 | 5817 | trace | 99.84 |
| $^{195}$Ir | 2.3 h | 0 | $^{195}$Pt | 29832 | <1 | 75.80 | 32648 | trace | 79.88 |

| Prod. | $T_{1/2}$ | Cool. (h) | Target | Fusion | | | Fission | | |
|---|---|---|---|---|---|---|---|---|---|
| | | | | $A_m$ | A | P | $A_m$ | A | P |
| | | | $^{198}Pt$ | 45500 | trace | 6.18 | 45301 | trace | 6.07 |
| | | | $^{193}Ir$ | trace | <1 | 0.02 | <0.1 | 218.2 | 1.17 |
| $^{197}Pt$ | 19.9 h | 24 | $^{198}Pt$ | <0.1 | 243.0 | 100.00 | trace | 1.2 | 99.74 |
| | | | $^{196}Pt$ | <0.1 | 88.3 | 46.81 | <0.1 | 170.6 | 99.84 |
| | | | $^{197}Au$ | 6 | <1 | 99.87 | 2 | trace | 99.96 |
| | | | $^{200}Hg$ | 5723 | <0.1 | 100.00 | 5714 | trace | 99.99 |
| $^{198}Au$ | 2.7 d | 24 | $^{198}Hg$ | 2 | <1 | 5.77 | 4 | trace | 14.13 |
| | | | $^{197}Au$ | <1 | 1694.0 | 82.71 | 7 | 37120.0 | 69.68 |
| $^{199}Au$ | 3.1 d | 24 | $^{199}Hg$ | 1446 | 1.1 | 98.26 | 1477 | trace | 99.35 |
| | | | $^{198}Pt$ | 214 | 193.3 | 99.90 | 1534 | 1870.0 | 100.00 |
| | | | $^{202}Hg$ | 1539 | trace | 99.98 | 1539 | trace | 99.92 |
| $^{209}Pb$ | 3.2 h | 0 | $^{209}Bi$ | 10576 | <1 | 99.00 | 8517 | trace | 99.59 |
| | | | $^{208}Pb$ | trace | <1 | <1 | trace | <1 | 48.70 |